\documentclass[
reprint,
superscriptaddress,
showpacs,
aps,
]{revtex4-1}

\usepackage{amsmath}
\usepackage{amssymb}%
\usepackage{natbib}
\usepackage{graphicx}
\usepackage{dcolumn}
\usepackage{bm}
\usepackage{ulem,color}

\newcommand{\uar}{\uparrow}
\newcommand{\dar}{\downarrow}
\newcommand{\lag}{\langle}
\newcommand{\rag}{\rangle}
\newcommand{\lt}{\left}
\newcommand{\rt}{\right}

\newcommand{\hspc}{\hspace{1em}}
\newcommand{\htab}{\hspace{2em}}

\newcommand{\mrm}{\mathrm}
\newcommand{\mcl}{\mathcal}

\newcommand{\mbf}{\mathbf}

\newcommand{\rnp}{\mrm{p}}
\newcommand{\rnm}{\mrm{m}}
\newcommand{\rnu}{\mrm{u}}
\newcommand{\rnl}{\mrm{l}}

\newcommand{\im}{\mrm{i}}

\begin{document}

\title{Magnetic Phase Diagram of Frustrated Spin Ladder}
 
\author{Takanori Sugimoto}
\email{sugimoto.takanori@rs.tus.ac.jp}
\affiliation{Department of Applied Physics, Tokyo University of Science, Katsushika, Tokyo 125-8585, Japan}
\author{Michiyasu Mori}
\affiliation{Advanced Science Research Center, Japan Atomic Energy Agency, Tokai, Ibaraki, 319-1195, Japan}
\author{Takami Tohyama}
\affiliation{Department of Applied Physics, Tokyo University of Science, Katsushika, Tokyo 125-8585, Japan}
\author{Sadamichi Maekawa}
\affiliation{Advanced Science Research Center, Japan Atomic Energy Agency, Tokai, Ibaraki, 319-1195, Japan}

\date{\today}

\begin{abstract}
Frustrated spin ladders show magnetization plateaux depending on the rung-exchange interaction and frustration defined by the ratio of first and second neighbor exchange interactions in each chain. 
This paper is the first report on its magnetic phase diagram. 
Using the variational matrix-product state method, we accurately determine phase boundaries. 
Several kinds of magnetization plateaux are induced by the frustration and the strong correlation among quasi-particles on a lattice. 	
The appropriate description of quasi-particles and their relevant interactions are changed by a magnetic field. 
We find that the frustration differentiates the triplet quasi-particle from the singlet one in kinetic energy. 
\end{abstract}


\maketitle

\section{Introduction}
Quantum phase transitions between spin liquid and magnetization-plateau (MP) phases are extensively studied in various low-dimensional spin systems with an interest of energy gaps emerging from spontaneously broken symmetry. 
These transitions are also the issue of correlated many-body systems and are closely related to the Mott transition in correlated electrons.  
According to the preceding studies~\cite{Lieb61,Haldane82,Oshikawa97,Totsuka98,Oshikawa03}, one of key factors is a commensurability energy of elementary excitations in spin systems. 
When the kinetic term is dominant as compared with the potential in an effective model, a spin liquid phase, i.e., metallic phase, is realized and its ground state is gapless. 
On the other hand, if the commensurability energy is relevant under a certain condition, the ground state can have a gap, and a MP phase, i.e., insulator phase, simultaneously appears. 
Oshikawa--Yamanaka--Affleck shows the condition for a magnetization per unit cell $M$ to have the MPs~\cite{Oshikawa97,Totsuka98}.
According to their argument, as increasing a commensurability energy (or potential energy) by introducing a perturbation competing against the kinetic energy, the system undergoes the gapless-to-gapped (spin-liquid-to-MP) transition, whose universality class is the same as the Berezinskii--Kosterlitz--Thouless (BKT) transition. 
To identify elementary excitations and to determine their effective models are the crucial points to study the spin-liquid-to-MP transition. 

Previous studies on the two-leg spin ladder (2LSL) show that a quasi-particle different from spinon plays the role of elementary excitation in small-magnetization ($M\sim 0$) and nearly-saturated ($M\sim M_{\mrm{sat}}$) regions in the strong rung limit~\cite{Dagotto96,Gopalan94,Sushkov98,Giamarchi99,Lavarelo12,Sugimoto15}.
The quasi-particle at $M\sim 0$ ($M\sim M_{\mrm{sat}}$) is formed by a triplet (singlet) state on a rung, and is called triplon (singlon) as a hard-core boson~\cite{Gopalan94}. 
The triplon (singlon) is a particle in a sea of rung singlets (triplets).
It is noted that triplon and singlon can be associated with particle and hole, respectively. 
Since the chemical potential of the hard-core boson corresponds to a magnetic field, a half-filled band of hard-core bosons can be realized at a certain magnetic field giving $M\sim M_{\mrm{sat}}/2$. 
In this condition, instead of hard-core bosons, i.e., triplon and singlon, a quasi-spin formed by the triplet and the singlet states well describes low-energy excitations around $M\sim M_{\mrm{sat}}/2$~\cite{Giamarchi99,Sugimoto15}. 
Thus, it is important to choose an appropriate quantum object to describe magnetic excitations, and in the 2LSL our choice is changed by a magnetic field, i.e., magnetization.  

Here, we switch on a frustration, which is given by a bond between second neighbor sites in each chain of two legs. Note that a bond diagonally running across a ladder plaquette is not considered in this paper.
The frustrated 2LSL (F-2LSL) has a gap in the ground state~\cite{Lavarelo12} and exhibits 1/3-, 1/2-, and 2/3-MP phases due to the competition between a frustration and the rung coupling~\cite{Sugimoto15}. 
Without rung coupling, we obtain two frustrated spin chains, which have a gap in their ground state and show the 1/3-MP phase.
It is explained by a boson-field model with a commensurability energy 
originating from the frustration~\cite{Haldane82,Majumdar69}.
On the other hand, bosons with multiple components derived from the spin ladder have some types of interactions, whose commensurability energy originates from inter-chain couplings~\cite{Totsuka98,Nersesyan98}. 
These mechanisms of commensurability thus are different between chain and ladder. 
In short, the F-2LSL has various aspects in its character, i.e., frustrated spin chain and frustrated spin ladder. 
Therefore, a key factor of those MP phases is not obvious.

In this paper, we show a magnetic phase diagram of the F-2LSL with respect to the frustration in each leg and the rung coupling. 
Using the variational matrix-product state (VMPS) method~\cite{schollwock11}, each phase boundary can be accurately determined. 
Differences between triplon and singlon and between the 1/3- and 2/3-MP phases is discussed. 
At the same time, triplon-singlon correspondence based on the particle-hole symmetry of quasi-particles, is justified in the strong rung limit.
Furthermore, characters of the quasi-particles in each phase are also shown. 

The contents of this paper are as follows.
In Sec.~II, we introduce the model Hamiltonian of a frustrated spin ladder, and also present an effective Hamiltonian in the strong rung limit.
Three MPs in our model are explained by the quasi-spin picture in an effective Hamiltonian~\cite{Sugimoto15}.
Our method to obtain the magnetic phase diagram and its condition of numerical calculations are mentioned in Sec.~III.
We show the magnetic phase diagram in Sec.~IV, and explain the origin of discrepancy between the 1/3- and 2/3-MP phase boundaries using quasi-particles in Sec.~V.
Summary is given in Sec.~VI.

\section{Model: Frustrated 2-Leg Spin Ladder}
We theoretically study magnetic phase diagram of F-2LSL given by,
\begin{equation}
\mcl{H}=\mcl{H}_\parallel+\mcl{H}_\perp+\mcl{H}_{Z},
\label{eq:ham0}
\end{equation}
with
\begin{eqnarray}
\mcl{H}_\parallel&=&\sum_{L=1,2} J_L 
\sum_{j=1}^N\sum_{i=\mrm{u,l}}\bm{S}_{j,i}\cdot\bm{S}_{j+L,i},\\
\mcl{H}_\perp&=&J_\perp\sum_{j=1}^N\bm{S}_{j,\mrm{u}}\cdot\bm{S}_{j,\mrm{l}}, 
\\
\mcl{H}_{Z}&=& -H \sum_{j=1}^N \sum_{i=\mrm{u,l}} {S}_{j,i}^z,
\end{eqnarray}
where $\bm{S}_{j,\rnu}$ ($\bm{S}_{j,\rnl}$) is the $S=\frac{1}{2}$ spin 
operator on $j$th rung in the upper (lower) leg.
There are three types of anti-ferromagnetic Heisenberg interactions: a 
nearest-neighbor coupling on a rung bond $J_\perp$, a nearest-neighbor 
coupling 
$J_1$ and a next-nearest-neighbor coupling $J_2$ in the leg direction.
For simplicity, we introduce two rational angles, leg-rung ratio 
$\alpha=\tan^{-1}(J_\parallel/J_\perp)$ and frustration 
$\beta=\tan^{-1}(J_2/J_1)$ with $J_\parallel=\sqrt{J_1^2+J_2^2}$.

It is useful to note several limits and schematic picture of the MP states as 
discussed in Refs.~\cite{Lavarelo12,Sugimoto15}.
In the limit of weak rung coupling, $J_\perp\ll J_1, J_2$ ($\alpha\sim 
\pi/2$), this model approaches two decoupled frustrated spin chains, while a 
non-frustrated spin ladder is obtained in another limit of weak frustration, 
$J_2 \ll J_1, J_\perp$ ($\beta \sim 0$).
This model (\ref{eq:ham0}) bridges between the frustrated spin chains and the 
non-frustrated spin ladder through $J_\perp$ and $J_2$.
This nature appears also in zero magnetic field as two different phases: 
columnar-dimer and rung-singlet phases.
In the columnar-dimer phase, the ground state is composed of two degenerated 
states with spontaneously-broken translational symmetry~\cite{Majumdar69}, 
while the rung-singlet phase has no degeneracy in its ground state.

The nature bridging two different systems is more apparent in the MP states 
with finite magnetizations.
For instance, the 1/2-MP phase is not allowed in the weak rung limit, because 
a frustrated spin chain does not exhibit a 1/2-MP phase.
However, it becomes possible in the strong rung limit with a frustration, 
because the effective Hamiltonian around 1/2-MP corresponds to a frustrated 
quasi-spin chain with an effective magnetic field given by~\cite{Sugimoto15},
\begin{equation}
\mcl{H}_{\mrm{eff}}^{(1)} = \mcl{P}\mcl{H}\mcl{P} = \mcl{H}_\parallel^\prime + 
\mcl{H}_Z^\prime \label{eq:ham1}
\end{equation}
with
\begin{eqnarray}
\mcl{H}_\parallel^\prime \! &=& \!\sum_{\eta=1,2}\sum_j \Bigg[ 
{J_{\eta}^z}^\prime T_{j}^zT_{j+\eta}^z + {J_{\eta}^x}^\prime 
(T_{j}^xT_{j+\eta}^x+T_{j}^yT_{j+\eta}^y )\Bigg], \\
\mcl{H}_Z^\prime &=& - H^\prime \sum_j T_{j}^z,
\end{eqnarray}
where $\mcl{P}$ is a projection operator, which projects out two of rung 
triplets as irrelevant high-energy states, and $\bm{T}_{j}$ is the quasi-spin 
operator at $j$th rung composed by singlet and triplet states on a rung (see 
Appendix~A for the detailed derivation).
The XY- and Ising-components of effective exchange interactions are denoted by 
${J_{\eta}^x}^\prime=J_{\eta}$ and ${J_{\eta}^z}^\prime=J_{\eta}/2$ with 
$\eta=1,2$, respectively. 
In this picture, relationship between the quasi-spin magnetization $M^\prime$ 
and the real magnetization $M$ is given by 
$M^\prime/M_{\mrm{sat}}^\prime=2M/M_{\mrm{sat}}-1$, so that the 1/2-MP state 
is 
regarded as the quasi-spin dimer state in the Majumdar-Gosh 
Hamiltonian~\cite{Majumdar69}.
Additionally, this picture in the strong rung limit derives other MPs, namely 
1/3- and 2/3-MP states, which correspond to $\mp 1/3$-MP states in terms of 
quasi-spins respectively, as discussed in several spin-$\frac{1}{2}$ 
chains~\cite{Hida94,Okunishi03}. 
In the weak rung limit, however, there does not appear 2/3-MP state, but only 
1/3-MP state does.
Therefore, we expect that a difference between the triplet and singlet states 
will be enhanced as decreasing the rung interaction from the strong rung 
limit. 
This is a kind of particle-hole symmetry breaking of hard-core quasi-particles.

\section{Variational Matrix-Product State Method}
To clarify phase boundaries of the MP states, we apply the VMPS method~\cite{schollwock11} to calculate the ground-state energies $E_M(N)$ with finite magnetization $M$ in finite system size $N$, which is the number of rungs.
In the VMPS method, we decompose a trial wave function 
$|\psi\rag=\sum_{\bm{\sigma}}\psi_{\bm{\sigma}}|\bm{\sigma}\rag$ into a matrix product form of open boundary condition: 
\begin{align}
\psi_{\bm{\sigma}}=\mbf{M}^{\sigma_1}\mbf{M}^{\sigma_{2}}\cdots\mbf{M}^{\sigma_{2N}},
\end{align}
where $\bm{\sigma}=\{\sigma_{i}\,|\, i=1,2,\cdots,2N\}$ is a set of the local spin indices with $\sigma_i=\uar,\dar$. 
The so-called canonical matrix is denoted by $\mbf{M}^{\sigma_i}$, whose elements $M_{n,n^\prime}^{\sigma_i}$ are given by local spin index $\sigma_i$ of $i$th site and auxiliary index $n$ ($n^\prime$) denoting the entangled state number in the left (right) side system. 
It should be noted that the wavefunction can be factorized by the local matrix, i.e., $\mbf{M}^{\sigma_i}$, thanks to the auxiliary index (see Appendix~B for more detail).
The Hamiltonian is also rewritten by a matrix-product operator:
\begin{equation}
\mcl{H}_\parallel+\mcl{H}_\perp=\mbf{H}_1\mbf{H}_2\cdots\mbf{H}_{2N}
\end{equation}
where local Hamiltonian $\mbf{H}_i$ is composed of only local spin operator $\bm{S}_i$.
In this form, we can deal the spin degrees of freedom site by site.
Actually, the spin indices can be contracted as an expectation value of the Hamiltonian except for a certain site, which we focused on.
After this contraction, we obtain the effective Hamiltonian for the site $\tilde{\mbf{H}}_i$, whose dimension equals to the square of the dimension of local matrix $\mbf{M}^{\sigma_i}$.
The variational calculation of the local matrix $\mbf{M}^{\sigma_i}$ is equivalent to an eigenvalue problem of the effective Hamiltonian $\tilde{\mbf{H}}_i$ with respect to the eigenvector $\mbf{M}^{\sigma_i}$.
Therefore, we can optimize the wave function by solving the eigenvalue problem site by site.
This approach is mathematically equivalent to the density-matrix renormalization-group method~\cite{white92}, which is one of most powerful methods for one-dimensional quantum systems.
This method appropriately deals with the quantum entanglement of different sites, so that we can accurately obtain the ground state and its energy.
Furthermore, it is also practically important that the VMPS method simplifies numerical coding and accelerates its development speed as compared with the density-matrix renormalization-group method.
Thus, we use this method to calculate the ground-state energy $E_M(N)$.

The MP gaps $\Delta_M$ are obtained as extrapolated values of $E_{M+1}(N)+E_{M-1}(N)-2E_M(N)$ with respect to inverse system size $1/N\to0$.
The ground-state energies $E_M(N)$ are calculated up to $N=144$ rungs with keeping the number of states $m=1000$ at most.
In this calculation, we confirm that the truncation error is less than about $10^{-5}$ (See also Appendix~C for details of numerical calculation).
The phase boundaries are determined as points at which the MP gap turns from positive to zero (or negative) with respect to the control parameter in Fig.~\ref{fig:pd}. 

\section{Magnetic Phase Diagram}
\begin{figure}[h]
\includegraphics[width=0.38\textwidth]{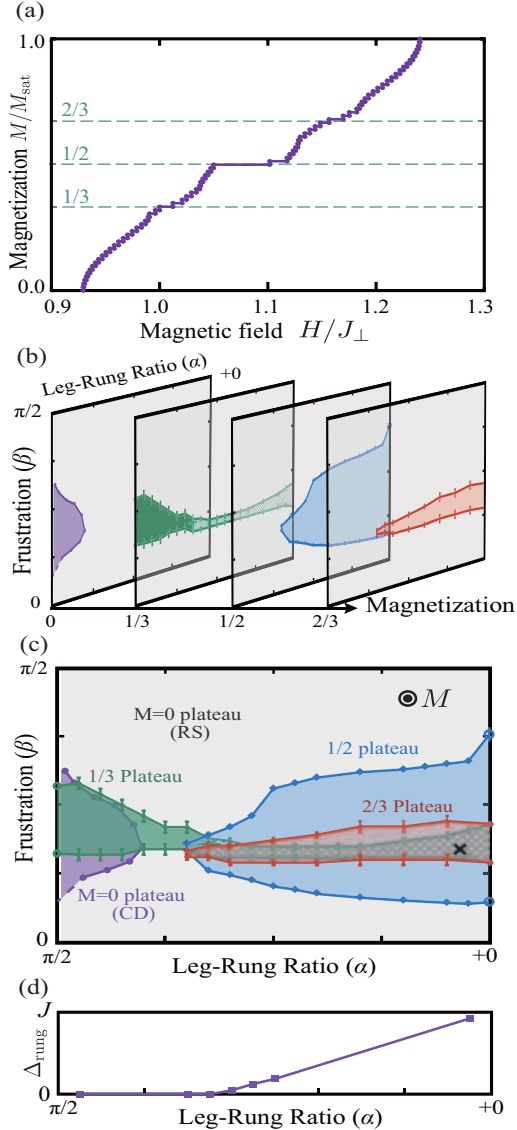}
\caption{(a) An example of magnetization curve with $\alpha\cong 0.037\pi$ and $\beta\cong 0.172\pi$, which are located at ``$\times$'' in (c). (b) Magnetic phase diagrams and (c) comparison of the 
phase diagrams of F-2LSL, where the boundary of $M=0$ plateau phases, columnar 
dimer (CD) and rung singlet (RS), is referenced by Ref.~\cite{Lavarelo12}. 
 In these diagrams, uncolored (gray) regions denote the gapless phases, except for $M=0$.
(d) Energy gap $\Delta_{\mrm{rung}}$ as a function of $\alpha$, which 
indicates a gap opening behavior at a critical point $\alpha_c\cong 0.350\pi$ 
in 1/3-magnetized ground states. This point corresponds to the boundary between 
the green and shaded green regions in Figs.~(a) and (b). Error bars are set 
under consideration of fluctuation energy estimated by the truncation error.
\label{fig:pd}}
\end{figure}

Figure~\ref{fig:pd} shows the magnetic phase diagram with several MP phases. 
Typical magnetization curve with three MPs, i.e., $M=$1/3, 1/2, and 2/3, is shown in Fig. 1(a) for $\alpha\cong 0.037\pi$ and $\beta=0.172\pi$ ($J_2/J_1=0.6$ and $J_1/J\perp=0.1$) ~\cite{Sugimoto15}. 
In Fig.~\ref{fig:pd}(b), possible regions of MP phases at each $M$ are 
shown with respect to rung interactions and frustration, which are parametrized 
by $\alpha$ and $\beta$, respectively. 
The four figures in Fig.~\ref{fig:pd}(b) are mapped into one plane as 
shown in Fig.~\ref{fig:pd}(c).
Firstly, we can confirm a coincidence of the phase boundaries of 1/3- and 
2/3-MP phases in the strong rung limit [$\alpha\to+0$ in Fig.~\ref{fig:pd}(c)].
This coincidence can be explained well in quasi-spin effective Hamiltonian, because these MP states correspond to negative and positive 1/3-MP states in terms of quasi-spins, i.e., the 1/3-MP state is equivalent to the 2/3-MP state with respect to the quasi-spin inversion.
The coincidence, however, is broken by a small intra-chain interaction, especially for the upper boundaries.
This discrepancy is more emphasized as increasing intra-chain interactions ($\alpha\to \pi/2$), and at last, 2/3-MP state disappears, though 1/3-MP state can survive.
The disappearance of 2/3-MP state is the consequence of the frustrated spin chain involved in the F-2LSL.
We can also confirm a boundary between chain and ladder by the rung parity $P_{\mrm{rung}}=\prod_j(2\bm{S}_{j,\mrm{u}}\cdot\bm{S}_{j,\mrm{l}}+\dfrac{1}{2})=\pm1$ and degeneracy of the ground states for 1/3-MP phase.
This is because the 1/3-MP state in the weak rung limit ($\alpha\to\pi/2$) is two frustrated spin chains, which is different from the strong rung limit ($\alpha\to+0$). 
The ground state of two frustrated spin chains has two-fold degeneracy in terms 
of $P_{\mrm{rung}}=\pm1$, while there is no degeneracy in the strong rung limit 
and the parity is given by $P_{\mrm{rung}}=(-1)^{2N/3}$ with 1/3 magnetization 
in a $N$-rung system where the triplet (singlet) state with the rung parity 
$+1$ ($-1$) occupies $N/3$ ($2N/3$) rungs. 
Figure~\ref{fig:pd}(d) shows a gap $\Delta_{\mrm{rung}}$, which is 
defined as an energy gap between the $P_{\mrm{rung}}=\pm1$ ground states with 
1/3 magnetization.
This is obtained by the VMPS method with an auxiliary field $\mu |P_{\mrm{rung}}\pm 1|$~\cite{note1}. 
We can see that the gap $\Delta_{\mrm{rung}}$ opens at $\alpha\sim 0.325\pi$ as increasing rung coupling (decreasing $\alpha$) at the 1/3-MP state. 
This boundary of two different 1/3-MP phases gives a good correspondence to the critical point ($\alpha\sim 0.350\pi$) where the 2/3-MP phase disappears as decreasing $\alpha$. 
Moreover, the critical point of 1/2-MP phases is interestingly close to the boundary, so that the boundary intuitively indicates a crossover between two different models, namely chain and ladder.

\section{Higher-Order Effective Hamiltonian}
Based on the discussion above, in the smaller $\alpha$ region than the boundary ($\alpha\lesssim\frac{3}{8}\pi$), the quasi-spin picture should work well.
To explain the discrepancy between the 1/3- and 2/3-MP phases in this region, we consider higher-order approximation of the original Hamiltonian as follows,
\begin{equation}
\mcl{H}_{\mrm{eff}}^{(2)} = \mcl{P} \mcl{V}(E_0-\mcl{H}_\perp-\mcl{H}_Z)^{-1} \mcl{Q}\mcl{H}_\parallel \mcl{P} = \mcl{H}_{\mrm{eff}}^{(2a)} + \mcl{H}_{\mrm{eff}}^{(2b)},
\end{equation}
where $\mcl{Q}$ is an orthocomplemental projection of $\mcl{P}$ and we use the unperturbed ground-state energy as an approximated eigen-energy (see Appedix~A).
As a second-order term of effective Hamiltonian, we find a symmetry-breaking term of quasi-spin inversion ($T_j^z\to-T_j^z$, $T_j^+\to T_j^-$, and its Hermite conjugate) given by,
\begin{eqnarray}
\mcl{H}_{\mrm{eff}}^{(2a)} &=& -\sum_j \Big[\sum_{L=1,2} J_L^{\prime\prime} (T_{j}^+ T_{j+2L}^- +T_{j}^- T_{j+2L}^+)T_{j+L}^z \notag\\
& & \hspace{1.5em}+ J^{\prime\prime} (T_{j}^+ T_{j+3}^- + T_{j}^- 
T_{j+3}^+)(T_{j+1}^z+T_{j+2}^z)\Big].\label{asymH}
\end{eqnarray}
The other term $\mcl{H}_{\mrm{eff}}^{(2b)}$ has the same form as the first-order Hamiltonian.
By the quasi-spin inversion, the Hamiltonian $\mcl{H}_{\mrm{eff}}^{(2a)}$ changes the sign.
In these terms, the second-neighbor hopping with an intermediate-site magnetization, i.e., the first term of $\mcl{H}_{\mrm{eff}}^{(2a)}$, can affect the second-neighbor hopping of the first-order Hamiltonian, so that this term can change the BKT transition point.
To clarify an effect of the first term in (\ref{asymH}), we consider the mean-field Hamiltonian given by,
\begin{equation}
H_{\mrm{MF}}
 \cong - J_1^{\prime\prime}\bar{T}^z \sum_j (T_{j}^+ T_{j+2}^- +T_{j}^- 
 T_{j+2}^+) - 2J_1^{\prime\prime} \chi_1 \sum_j T_{j}^z,
\end{equation}
where $\bar{T}^z=\sum_j \lag T_j^z\rag/N$ and $\chi_1=\sum_j \Re \lag T_{j-1}^+ T_{j+1}^- \rag/N$.
The second term changes the critical magnetic field, at which the MP state appears.
Since $\bar{T}^z$ changes the sign from negative to positive at the half filling, the first term turns from additive to subtractive to the second-neighbor hopping, as increasing the magnetization.
In particular, the second-neighbor interaction dominates around the upper boundaries of the 1/3- and 2/3-MP states as compared with the first-neighbor interaction.
Here, a large hopping interaction suppresses the MP gap and thus, this term affects the MP boundaries as follows: the upper boundary of the 1/3-MP phase slides down with a help of this term, but that of the 2/3 slides up.

\begin{figure}[htb]
\includegraphics[width=0.45\textwidth]{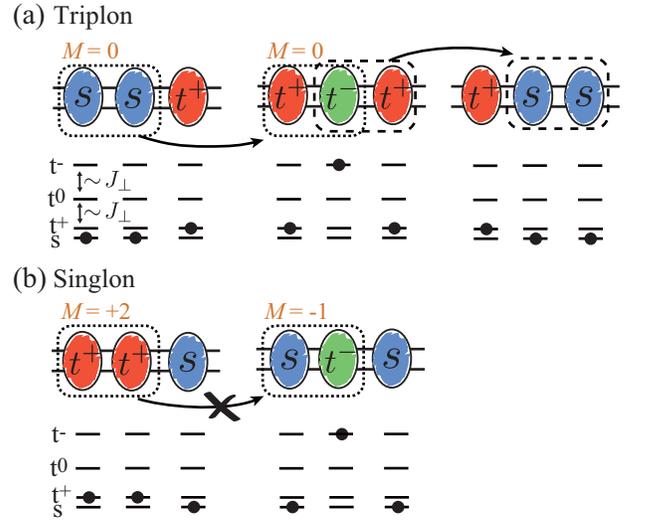}
\caption{Schematic difference of the second-order term of the effective 
Hamiltonian (\ref{asymH}) between (a) triplon and (b) singlon.
Colored ovals including a character $s$ ($t^\pm$) denote the singlet (triplet) states of rung.
The energy levels of each rung, i.e., the singlet $s$ and triplets $t^{\pm,0}$, are shown as solid lines below the oval, and the occupied state is represented by a black dot. 
In the small-magnetization (nearly-saturated) region with a strong rung interaction, the triplet (singlet) state behaves as a hard-core boson, which is called triplon (singlon), in the sea of singlets (triplets). 
A perturbative leg interaction does allow pair annihilation (dotted-lined area) and creation (dashed-lined area) of singlet in (a), but prohibit those of triplet in (b), because of the conservation law of magnetization $M$.
In (a), we can see triplon hopping to the 2nd neighboring rung, through the intermediate state with an occupation of high energy levels.
\label{fig:st}}
\end{figure} 

The quasi-spin picture in a small magnetization region and a nearly-saturated regions can be associated with triplon and singlon, respectively.
From the quasi-particle point of view, the asymmetric term results in difference of kinetic energy (see Fig.~\ref{fig:st}).
In small-magnetization region [Fig.~\ref{fig:st}(a)], a small leg interaction can create and annihilate a singlet pair by using a neighboring triplet pair of opposite magnetizations $M=\pm 1$ as keeping the conservation law of magnetization, where triplet with negative magnetization ($t^-$) has a much higher energy than those of a triplet state ($t^+$) and a singlet state ($s$) and plays the role of an intermediate state.  
On the other hand, in nearly-saturated region [Fig.~\ref{fig:st}(b)], a leg perturbation cannot create and annihilate a triplet pair of the same magnetization $M=+1$ by using any other two rung states, because two rung states with the magnetization $M=+2$ are unique. 

In fact, supposed the non-interacting case, higher-order long-ranged 
hopping terms 
renormalize the velocity of hard-core boson,
$v_0= \partial \epsilon_k/\partial k\sim k\sum_{L=1,2,\cdots} L^2t_L =t^\prime 
k$ around the minimum energy in the dispersion relation 
$\epsilon_k=-\sum_Lt_L\cos(Lk)$, where $t_L$ is $L$th neighbor hopping obtained 
as higher-order approximation of $J^x$s and $t^\prime$ is a renormalized 
hopping.
Thus, the kinetic energy of quasi-particle in small-magnetization region is relatively larger than that in nearly-saturated region.

\section{Conclusion}
In summary, we determined the magnetic phase diagram of the frustrated two-leg spin ladder by using the VMPS method.
We found the 1/3-, 1/2-, and 2/3-MP phases in the diagram with respect to the frustration and the inter-chain interaction. 
All MPs are suppressed around one point ($\alpha\sim \frac{3}{8}\pi$ and $\beta\sim\frac{3}{16}\pi$) in the phase diagram. 
This shows crossovers between chain and ladder, and between frustrated and not frustrated systems.   
Furthermore, the difference between singlet- and triplet-based quasi-particles is clarified even with a small leg interaction.
This difference originates from the second-order perturbation of leg interaction, which implies the kinetic energy of quasi-particle in small-magnetization region is larger than that in nearly-saturated region.
The quasi-particles, triplon and singlon, are elementally excitations not only in ladder systems such as BiCu$_2$PO$_6$~\cite{Tsirlin10,Kohama12,splinter16,hwang16,Plumb16,note2}, but also strong dimer models such as two-dimensional Shastry-Sutherland compound SrCu$_2$(BO$_3$)$_2$~\cite{Kodama02} and three-dimensional spin-dimer compound TlCuCl$_3$~\cite{Ruegg03,kimura16}.
This implies that the magnetic phase diagram is the starting point to search for multiferroic materials.
Our results will be useful also for spin transport and its application such as spintronics\cite{uchida08,adachi13,hirobe16}.

\begin{acknowledgments}
We would like to thank M. Fujita and O. P. Sushkov for valuable discussions. 
This work was partly supported by Grant-in-Aid for Scientific Research (Grant No.25287094, No.26103006, No.26108716, No.26247063, No.26287079, No.15H03553, and No.15K05192), Grants-in-Aids for Young Scientists (B) (Grant No.16K17753), the CDMSI project on a post-K computer, and the inter-university cooperative research program of IMR, Tohoku University.
Numerical computation in this work was carried out on the supercomputers at JAEA and the Supercomputer Center at Institute for Solid State Physics, University of Tokyo.
\end{acknowledgments}

\appendix
\section{Derivation of second-order terms of effective Hamiltonian}

In order to obtain the effective Hamiltonian of quasi-spins step by step, we start from the original spin Hamiltonian given by
\begin{equation}
\mcl{H}=\mcl{H}_{\parallel}+\mcl{H}_{\perp}+\mcl{H}_{Z} \label{eq:ham}
\end{equation}
with
\begin{align}
\mcl{H}_{\parallel}&=\sum_{L=1,2}\mcl{H}_{L}=\sum_{L=1,2}J_L\sum_{i=u,l}\sum_{j} \bm{S}_{j,i} \cdot \bm{S}_{j+L,i}, \\
\mcl{H}_{\perp}&=J_\perp\sum_j \bm{S}_{j,\rnu} \cdot \bm{S}_{j,\rnl}, \hspc \mcl{H}_{Z}=-H^z\sum_{i=u,l}\sum_j S_{j,i}^z.
\end{align}

In the following, we consider the strong rung-coupling limit, $J_1/J_\perp \ll 1$ and $J_2/J_\perp\ll 1$, so that the leg Hamiltonian $\mcl{H}_{\parallel}$ is dealed as a perturbation in basis diagonalizing the rung Hamiltonian $\mcl{H}_{\perp}+\mcl{H}_{Z}$.

\subsection{Bond-operator transform}
The diagonalization of the rung Hamiltonian is obtained by using the bond-operator representation as follows,
\begin{align}
T_{j,\rnp}^\dagger&=\frac{\im}{\sqrt{2}} \Big\{ S_{j,\rnu}^+ \exp\lt[\frac{\im\pi}{2} \lt(S_{j,\rnl}^z+\frac{1}{2}\rt)\rt] \notag\\
&\hspc\htab + S_{j,\rnl}^+ \exp\lt[-\frac{\im\pi}{2}\lt(S_{j,\rnu}^z+\frac{1}{2}\rt) \rt] \Big\}, \\
m_{j,\rnp} &= T_{j,\rnp}^\dagger T_{j,\rnp} = \bm{S}_{j,\rnu}\cdot\bm{S}_{j,\rnl}-\lt(S_{j,\rnu}^z-\frac{1}{2}\rt)\lt(S_{j,\rnl}^z-\frac{1}{2}\rt)+3/4
\end{align}
Here, the creation and annihilation operators, $T_{j,\rnp}^\dagger$ and $T_{j,\rnp}$ obey on-site anti-commutation relation $\{T_{j,\rnp},T_{j,\rnp}^\dagger\}=1$ and commutation relation between different sites $[T_{j,\rnp},T_{k,\rnp}^\dagger]=[T_{j,\rnp}^\dagger,T_{k,\rnp}^\dagger]=[T_{j,\rnp},T_{k,\rnp}]=0$, and thus $m_{j,\rnp}$ is regarded as the number operator of the hard-core boson $T_{j,\rnp}$.

A dual operator of the hard-core boson is obtained in the same manner,
\begin{align}
T_{j,\rnm}^\dagger&= \frac{1}{\sqrt{2}} \Big\{S_{\rnu}^- \exp\lt[\frac{\im\pi}{2} \lt(S_{j,\rnl}^z+\frac{1}{2}\rt)\rt] \notag\\
&\hspc\htab- S_{\rnl}^- \exp\lt[-\frac{\im\pi}{2} \lt(S_{j,\rnu}^z+\frac{1}{2}\rt)\rt] \Big\}, \\
m_{j,\rnm} &= T_{j,\rnm}^\dagger T_{j,\rnm} = \bm{S}_{j,\rnu}\cdot\bm{S}_{j,\rnl}-\lt(S_{j,\rnu}^z+\frac{1}{2}\rt)\lt(S_{j,\rnl}^z+\frac{1}{2}\rt)+3/4
\end{align}
This operator also obeys the hard-core bosonic commutation relation. 
We note that these operators commute each other, so that the transform is regarded as that from hard-core boson to another hard-core boson.
Since spin-$\dfrac{1}{2}$ operator, namely the Pauli operator, obeys the hard-core bosonic commutation relation, this transform corresponds to that from real spins to quasi-spins.
This is simply explained by using four rung states of singlet and triplet: $|\mrm{s}\rag_j$, $|\mrm{t}^\alpha\rag_j$ where $\alpha=\pm ,0$.
The bond operator or its dual is rewritten by
\begin{align}
T_{j,\rnp}^\dagger &= |\mrm{t}^+\rag\lag \mrm{s}|+\im|\mrm{t}^0\rag\lag t^-|, \ m_{j,\rnp} = |\mrm{t}^+\rag\lag \mrm{t}^+|+|\mrm{t}^0\rag\lag \mrm{t}^0|,
\end{align}
or
\begin{align}
T_{j,\rnm}^\dagger &= |\mrm{t}^-\rag_j\lag \mrm{s}|_j+\im|\mrm{t}^0\rag_j\lag \mrm{t}^+|_j, \ m_{j,\rnm} = |\mrm{t}^-\rag_j\lag \mrm{t}^-|_j+|\mrm{t}^0\rag_j\lag \mrm{t}^0|_j.
\end{align}
This representation obviously leads to the hard-core bosonic commutation relation, and gives us a simple explanation of their role: $T_{j,\rnp}^\dagger$ ($T_{j,\rnm}^\dagger$) is an increase (decrease) operator of {\it magnetization}.

In fact, the magnetization on a rung reads
\begin{align}
m_{j,\rnp}-m_{j,\rnm}&=\lt(S_{j,\rnu}^z+\frac{1}{2}\rt)\lt(S_{j,\rnl}^z+\frac{1}{2}\rt)-\lt(S_{j,\rnu}^z-\frac{1}{2}\rt)\lt(S_{j,\rnl}^z-\frac{1}{2}\rt)\notag\\
&=S_{j,\rnu}^z+S_{j,\rnl}^z,  
\end{align}
and the rung interaction is given by
\begin{align}
\bm{S}_{j,\rnu}\cdot\bm{S}_{j,\rnl}=-(m_{j,\rnp}-1)(m_{j,\rnm}-1)+\frac{1}{4}.
\end{align}
Therefore, we obtain diagonalization of the rung Hamiltonian as a starting point,
\begin{widetext}
\begin{align}
\mcl{H}_{\perp}+\mcl{H}_{Z}&=\sum_j \lt\{J_\perp \lt(m_{j,\rnp}+m_{j,\rnm}-m_{j,\rnp}m_{j,\rnm}-\frac{3}{4}\rt)-H^z (m_{j,\rnp}-m_{j,\rnm})\rt\} \\
&=-\frac{3}{4}J_\perp N + \sum_j \lt\{[J_\perp(1-m_{j,\rnp}) +H^z] m_{j,\rnm} + (J_\perp-H^z) m_{j,\rnp} \rt\}\label{eq:rung}
\end{align}

The leg ineractions $\mcl{H}_\parallel=\sum_{L=1,2}\mcl{H}_{L}$ is also rewritten by, 
\begin{align}
\mcl{H}_{L}/J_L &= \sum_{i=\mrm{u,l}} \lt[S_{j,i}^z S_{j+L,i}^z + \frac{1}{2} (S_{j,i}^+ S_{j+L,i}^- + S_{j,i}^- S_{j+L,i}^+)\rt] =\mcl{H}_{L,\mrm{R}}+\mcl{H}_{L,\mrm{I}}+\mcl{H}_{L,\mrm{K}},
\end{align}
with
\begin{align}
\mcl{H}_{L,\mrm{R}} &= \frac{J_L}{2} \sum_{j}  (m_{j,\rnp} -m_{j,\rnm}) (m_{j+L,\rnp} -m_{j+L,\rnm}), \\
\mcl{H}_{L,\mrm{I}} &= -J_L \sum_{j} (T_{j,\rnp}^\dagger T_{j,\rnm}^\dagger - T_{j,\rnp}T_{j,\rnm})(T_{j+L,\rnp}^\dagger T_{j+L,\rnm}^\dagger - T_{j+L,\rnp}T_{j+L,\rnm}), \\
\mcl{H}_{L,\mrm{K}} &= \frac{J_L}{2} \sum_{j} \bigg\{T_{j,\rnp}^\dagger T_{j+L,\rnp} \cos\lt[\frac{\pi}{2} (m_{j,\rnm}-m_{j+L,\rnm})\rt] +T_{j,\rnm} T_{j+L,\rnm}^\dagger \cos\lt[\frac{\pi}{2} (m_{j,\rnp}-m_{j+L,\rnp})\rt] \notag\\
&\htab \hspace{1em} - T_{j,\rnp}^\dagger T_{j+L,\rnm}^\dagger \cos\lt[\frac{\pi}{2} (m_{j,\rnm}-m_{j+L,\rnp})\rt] -T_{j,\rnm} T_{j+L,\rnp} \cos\lt[\frac{\pi}{2} (m_{j,\rnp}-m_{j+L,\rnm})\rt]+\mrm{H.c.}\bigg\}.
\end{align}
\end{widetext}
The first term $\mcl{H}_{L,\mrm{R}}$ plays the role of magnetic repulsion between $L$th neighboring rungs.
The second term $\mcl{H}_{L,\mrm{I}}$ generates a coupling between singlet and $M=0$ triplet, namely $|\mrm{s}\rag_j$, $|\mrm{t}^0\rag_j$, which corresponds to a kinetic term of $M=0$ mode of triplon because a pair creation and annihilation can be diagonalized by using so-called Bogoliubov transform.
This term is negligible with large magnetic field in the strong rung-coupling limit, though it can emerge in a higher-order effective model as a two-body interaction of the hard-core bosons.
The third term represents a kinetic term of the hard-core boson with a cosine phase.
This term plays a key role in inducing a symmetry breaking of quasi-spin inversion.

\subsection{Projection into low-energy states}
We introduce the projection operator $\mcl{P}$ into a subspace based on low-energy states, and that into its orthocomplement subspace $\mcl{Q}$, i.e., $\mcl{P}+\mcl{Q}=1$~\cite{Kuramoto00}.

Suppose the eigen-equation of the original Hamiltonian $\mcl{H}$ is given by $\mcl{H}\psi=E\psi$, an effective Hamiltonian $\mcl{H}_{\mrm{eff}}$ is expected to satisfy
\begin{equation}
\mcl{H}_{\mrm{eff}} \mcl{P}\psi=E\mcl{P}\psi.
\end{equation}
Here, we divide the original Hamiltonian into a commutative Hamiltonian $\mcl{H}_0$ and non-commutative one $\mcl{V}=\mcl{H}-\mcl{H}_0$ with respect to $\mcl{P}$, $[\mcl{P},\mcl{H}_0]=0$.
If we take care of this relation
\begin{equation}
\mcl{Q}\mcl{V}\psi=(1-\mcl{P})(\mcl{H}-\mcl{H}_0)\psi=(1-\mcl{P})(E-\mcl{H}_0)\psi=(E-\mcl{H}_0)\mcl{Q}\psi,
\end{equation}
we obtain the following equation,
\begin{align}
\mcl{Q}\psi &=\mcl{Q}(E-\mcl{H}_0)^{-1}\mcl{V}\psi = \mcl{X}(\mcl{Q}+\mcl{P})\psi \notag\\
&= \mcl{X}\mcl{P}\psi + \mcl{X}(\mcl{X}\mcl{P}\psi+\mcl{X}\mcl{Q}\psi) = \cdots = \sum_{n=1}^\infty \mcl{X}^n \mcl{P} \psi 
\end{align}
where $\mcl{X}=\mcl{Q}(E-\mcl{H}_0)^{-1}\mcl{V}$.
Since the eigen-function is $\psi=(\mcl{P}+\mcl{Q})\psi=\sum_{n=0}^\infty \mcl{X}^n \mcl{P} \psi $, the effective Hamiltonian can obey the expected equation $\mcl{H}_{\mrm{eff}} \mcl{P}\psi=E\mcl{P}\psi$ with
\begin{equation}
\mcl{H}_{\mrm{eff}} = \mcl{P} \mcl{H} \sum_{n=0}^\infty \mcl{X}^n \mcl{P} = \mcl{P} \mcl{H} \lt[(E-\mcl{H}_0)^{-1} \mcl{Q}\mcl{V}\rt]^n \mcl{P}.
\end{equation}
The $n$th-order term of effective Hamiltonian is given by,
\begin{align}
\mcl{H}_{\mrm{eff}}^{(n)} &= \mcl{P} \mcl{H} \mcl{X}^n \mcl{P} = \mcl{P} (\mcl{H}_0+\mcl{V}) \lt[(E-\mcl{H}_0)^{-1} \mcl{Q}\mcl{V}\rt]^n \mcl{P}\notag\\
&= \mcl{P} \mcl{V} \lt[(E-\mcl{H}_0)^{-1} \mcl{Q}\mcl{V}\rt]^n \mcl{P},
\end{align}
Here, note that $\mcl{P}\mcl{Q}=0$.

In this article, we choose $\mcl{P}=\prod_j (1-m_{j,\rnm})$ with
\begin{align}
\mcl{H}_0 = \mcl{H}_{\perp}+\mcl{H}_{Z} +\sum_{L=1,2} \mcl{H}_{L,\mrm{R}}, \hspc \mcl{V} = \sum_{L=1,2} \mcl{H}_{L,\mrm{I}} + \mcl{H}_{L,\mrm{K}}.
\end{align}
This is appropriate for $J_\perp \sim H^z \gg |J_\perp - H^z| \sim J_1 \sim J_2$.
The reason is as follows.
In the unperturbed Hamiltonian Eq.~(\ref{eq:rung}), because of $\lag 1-m_{j,\rnp}\rag >0$, the chemical potential of $m_{j,\rnm}$ is much larger than that of $m_{j,\rnp}$, i.e., $J_\perp\lag 1-m_{j,\rnp}\rag +H^z \gg J_\perp-H^z$.
Therefore, since $\lag m_{j,\rnp}\rag \gg \lag m_{j,\rnm}\rag \sim 0$ at the ground state, the low-energy physics should be discussed in $m_{j,\rnm}=0$ states.

\begin{enumerate}
\item{Zero-th order}\\
By using the projection operator, we first obtain the zero-th order term as follows,
\begin{equation}
\mcl{H}_{\mrm{eff}}^{(0)} = \mcl{P}\mcl{H}_0\mcl{P} = E_0 + \mcl{H}_{\mrm{CP}}^{(0)} + \mcl{H}_{\mrm{R}}^{(0)}
\end{equation}
with
\begin{align}
&E_0 = -\frac{3}{4}J_\perp N, \hspc \mcl{H}_{\mrm{CP}}^{(0)} = (J_\perp-H^z)\sum_j m_{j,\rnp}, \notag\\
&\mcl{H}_{\mrm{R}}^{(0)} = \sum_{L=1,2} \frac{J_L}{2} \sum_j  m_{j,\rnp}  m_{j+L,\rnp} 
\end{align}
Here, we note $m_{k,\rnm}(1-m_{k,\rnm})=0$, and $N$ denotes the number of rungs.

\item{First order}\\
In the same manner, we obtain the first-order term given by,
\begin{equation}
\mcl{H}_{\mrm{eff}}^{(1)} = \mcl{P}\mcl{V}\mcl{P} = \mcl{H}_{\mrm{K}}^{(1)}
\end{equation}
with
\begin{align}
\mcl{H}_{\mrm{K}} = \sum_{L=1,2} \frac{J_L}{2} \sum_{j} (T_{j,\rnp} T_{j+L,\rnp}^\dagger +T_{j,\rnp}^\dagger T_{j+L,\rnp}).
\end{align}

The symmetric Hamiltonian with respect to quasi-spin inversion, namely particle-hole symmetry of the hard-core boson, corresponds to sum of these terms,
\begin{equation}
\mcl{H}_{\mrm{eff}}^{(0)}+\mcl{H}_{\mrm{eff}}^{(1)} = \mcl{H}_{\mrm{K}}^{(1)} + \mcl{H}_{\mrm{V}}^{(0)} + \mcl{H}_{\mrm{Z}}^{(0)} + \mrm{const.} \label{eq:ham1}
\end{equation}
with
\begin{align}
\mcl{H}_{\mrm{K}}^{(1)} &= \sum_{L=1,2} \frac{J_L}{2} \sum_{j} (T_{j}^+ T_{j+L}^- +T_{j}^- T_{j+L}^+) \\
\mcl{H}_{\mrm{V}}^{(0)} &= \sum_{L=1,2} \frac{J_L}{2} \sum_j  T_{j}^z  T_{j+L}^z, \\
\mcl{H}_{\mrm{Z}}^{(0)} &= \lt(J_\perp-H^z+\sum_{L=1,2}\frac{J_L}{2}\rt)\sum_j T_{j}^z
\end{align}
where we use the transform from the hard-core boson to quasi-spin, $T_{j,\rnp}^\dagger\to T_{j}^+$, $T_{j,\rnp}\to T_{j+L}^-$, and $m_{j,\rnp}\to T_j^z+\dfrac{1}{2}$.

\begin{widetext}
\item{Second order}\\
Finally, we show the second-order term as follows,
\begin{align}
\mcl{H}_{\mrm{eff}}^{(2)} &= \mcl{P} \mcl{V}(E-\mcl{H}_0)^{-1} \mcl{Q}\mcl{V} \mcl{P} \notag\\
&\cong \sum_j \sum_{L=1,2}  \Bigg\{ \lt(\frac{J_L^2}{2H^z} -\frac{J_L^2}{2(J_\perp+H^z)}\rt)(m_{j,\rnp}-1) (m_{j+L,\rnp}-1) \notag\\
&-\frac{J_L^2}{4(J_\perp+H^z)} (T_{j,\rnp} T_{j+2L,\rnp}^\dagger +\mrm{H.c.})(m_{j+L,\rnp}-1)  -\frac{J_1J_2}{4(J_\perp+H^z)} (T_{j,\rnp} T_{j+3,\rnp}^\dagger +\mrm{H.c.})(m_{j+L,\rnp}-1) \Bigg\}
\end{align}
In terms of quasi-spins, this term reads,
\begin{align}
\mcl{H}_{\mrm{eff}}^{(2)} \cong \mcl{H}_{\mrm{K}}^{(2)}+\mcl{H}_{\mrm{V}}^{(2)}+\mcl{H}_{\mrm{Z}}^{(2)}+\mcl{H}_{\mrm{A}}^{(2)}+\mrm{const.}
\end{align}
with
\begin{align}
\mcl{H}_{\mrm{K}}^{(2)} &= \sum_j \lt[\sum_{L=1,2} \frac{J_L^2}{8(J_\perp+H^z)} (T_{j}^+ T_{j+2L}^- +T_{j}^- T_{j+2L}^+)  +\frac{J_1J_2}{4(J_\perp+H^z)} (T_{j}^+ T_{j+3}^- + T_{j}^- T_{j+3}^+)\rt] \notag \\
\mcl{H}_{\mrm{V}}^{(2)} &= \sum_j \sum_{L=1,2} \lt(\frac{J_L^2}{2H^z} -\frac{J_L^2}{2(J_\perp+H^z)}\rt) T_{j}^z T_{j+L}^z, \hspc \mcl{H}_{\mrm{Z}}^{(2)} = -\sum_j \sum_{L=1,2} \lt(\frac{J_L^2}{2H^z} -\frac{J_L^2}{2(J_\perp+H^z)}\rt) T_{j}^z \notag \\
\mcl{H}_{\mrm{A}}^{(2)} &= -\sum_j \lt[\sum_{L=1,2} \frac{J_L^2}{8(J_\perp+H^z)} (T_{j}^+ T_{j+2L}^- +T_{j}^- T_{j+2L}^+)T_{j+L}^z  +\frac{J_1J_2}{4(J_\perp+H^z)} (T_{j}^+ T_{j+3}^- + T_{j}^- T_{j+3}^+)(T_{j+1}^z+T_{j+2}^z) \rt] \notag
\end{align}
In this Hamiltonian, the symmetry-breaking term emerges as $\mcl{H}_{\mrm{eff}}^{(2a)}\equiv\mcl{H}_{\mrm{A}}^{(2)}$.
Effects of this term are discussed in the main text. 
\end{widetext}
\end{enumerate}

\section{Brief Review of Variational Matrix-Product State Method}
In this section, we briefly review the VMPS method~\cite{schollwock11}.
In the VMPS method, we use mixed-canonical matrix-product state as a trial wave function:
\begin{align}
|\psi\rag&=\sum_{\bm{\sigma}}\psi_{\bm{\sigma}}|\bm{\sigma}\rag\notag\\
&=\sum_{\bm{\sigma}}\bm{A}^{\sigma_1}\mbf{A}^{\sigma_{2}}\cdots\mbf{A}^{\sigma_{m-1}}\mbf{X}^{\sigma_m}\mbf{B}^{\sigma_{m+1}}\cdots\mbf{B}^{\sigma_{2N-1}}(\bm{B}^{\sigma_{2N}})^T|\bm{\sigma}\rag,
\end{align}
where $\bm{\sigma}=\{\sigma_{i}\,|\, i=1,2,\cdots,2N\}$ is a set of the local spin indeces with $\sigma_i=\uar,\dar.$, and $\mbf{A}^{\sigma}$ ($\mbf{B}^{\sigma}$) represents the left- (right-) canonical matrices.
Here, the left- (right-) {\it canonical} matrix is defined by the contraction rules: $\sum_\sigma(\mbf{A}^{\sigma})^\dag\mbf{A}^{\sigma}=\mbf{1}$ and $\mbf{A}^{\sigma}(\mbf{A}^{\sigma^\prime})^\dag=\delta_{\sigma,\sigma^\prime}\cdot\mbf{1}$ ($\sum_\sigma\mbf{B}^{\sigma}(\mbf{B}^{\sigma})^\dag=\mbf{1}$ and $(\mbf{B}^{\sigma})^\dag\mbf{B}^{\sigma^\prime}=\delta_{\sigma,\sigma^\prime}\cdot\mbf{1}$) except for edge {\it vectors} $\bm{A}^{\sigma_1}$ and $\bm{B}^{\sigma_{2N}}$.
The edge vectors are given by $\bm{A}^{\uar}=\bm{B}^{\uar}=\begin{pmatrix}1 & 0\end{pmatrix}$ and $\bm{A}^{\dar}=\bm{B}^{\dar}=\begin{pmatrix}0 & 1\end{pmatrix}$.
We optimize the center trial matrix $\mbf{X}^{\sigma_m}$ using the variational approach with Lagrange multiplier $\epsilon$:
\begin{equation}
\frac{\delta}{\delta (\mbf{X}^{\sigma_m})^\dag} (\lag \psi|\mcl{H} |\psi\rag - \epsilon \lag\psi|\psi\rag) = \mbf{0}.
\end{equation}
This equation is rewritten by the following eigenvalue equation for the matrix $\mbf{X}^{\sigma_m}$,
\begin{equation}
\sum_{\sigma_m^\prime,i^\prime,j^\prime}\tilde{H}_{(i,j),(i^\prime,j^\prime)}^{\sigma_m,\sigma_m^\prime}X_{i^\prime,j^\prime}^{\sigma_m^\prime}=\epsilon \,X_{i,j}^{\sigma_m}.
\end{equation}
The reduced Hamiltonian $\tilde{\mbf{H}}^{\sigma_m,\sigma_m^\prime}$ is obtained by the contraction
\begin{align}
\tilde{\mbf{H}}^{\sigma_m,\sigma_m^\prime} = &\sum_{\bm{\sigma}\setminus\{\sigma_m\}}\sum_{\bm{\sigma}^\prime\setminus\{\sigma_m^\prime\}} (L_i^{\sigma_1,\sigma_2,\cdots,\sigma_{m-1}} R_j^{\sigma_1,\sigma_2,\cdots,\sigma_{m-1}})^\ast \notag\\
&\times H^{\bm{\sigma},\bm{\sigma}^\prime} L_{i^\prime}^{\sigma_1^\prime,\sigma_2^\prime,\cdots,\sigma_{m-1}^\prime} R_{j^\prime}^{\sigma_1^\prime,\sigma_2^\prime,\cdots,\sigma_{m-1}^\prime}
\end{align}
with matrix elements of the Hamiltonian $H^{\bm{\sigma},\bm{\sigma}^\prime}=\lag\bm{\sigma}|(\mcl{H}_\parallel+\mcl{H}_\perp)|\bm{\sigma}^\prime\rag$, and reduced left and right vectors 
\begin{align}
\bm{L}^{\sigma_1,\sigma_2,\cdots,\sigma_{m-1}}=\bm{A}^{\sigma_1}\mbf{A}^{\sigma_{2}}\cdots\mbf{A}^{\sigma_{m-1}},\\
\bm{R}^{\sigma_1,\sigma_2,\cdots,\sigma_{m-1}}=\mbf{B}^{\sigma_{m+1}}\cdots\mbf{B}^{\sigma_{2N-1}}\bm{B}^{\sigma_{2N}}.  
\end{align}
The reduced Hamiltonian can be obtained if we have the left- and right-canonical matrices, $\mbf{A}^{\sigma}$ and $\mbf{B}^{\sigma}$.
Therefore, the eigenvalue equation gives the optimized matrix $\mbf{X}^{\sigma_m}$ with fixed $\mbf{A}^{\sigma}$ and $\mbf{B}^{\sigma}$.
After the optimization, $\mbf{X}^{\sigma_m}$ can be decomposed into three matrices using the singular value decomposition:
\begin{equation}
\mbf{X}^{\sigma_m}=\mbf{A}^{\sigma_m}\mbf{\Lambda}\mbf{V}^\dag=\mbf{U}\mbf{\Lambda}\mbf{B}^{\sigma_m}
\end{equation}
where $\mbf{V}$ and $\mbf{U}$ are unitary matrices, and $\mbf{\Lambda}$ is a rectangular matrix which has singular values as diagonal elements. 
Here, the number of kept state $m$ is determined by the maximal number of kept diagonal elements. 
At the next step, the trial matrix is given by $\mbf{X}^{\sigma_{m+1}}=\mbf{\Lambda}\mbf{V}^\dag\mbf{B}^{\sigma_{m+1}}$ ($\mbf{X}^{\sigma_{m-1}}=\mbf{A}^{\sigma_{m-1}}\mbf{U}\mbf{\Lambda}$) during the left (right) sweep.
We repeat this sweep several times for all matrices while the convergence of energy $\epsilon$ is not an expected one.

\begin{figure}[hbt]
\begin{center}
\includegraphics[keepaspectratio,width=0.45\textwidth]{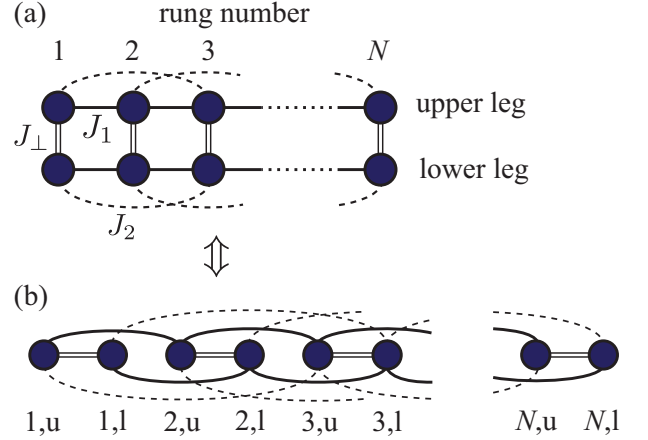}
	\caption{(a) Connectivity of the frustrated spin ladder. Balls represent the spins. Solid (dased) lines denote the 1st (2nd) neighbor exchange interactions in a leg. Double solid lines are the rung exchange interactions. (b) One-dimensional representation of the frustrated spin ladder. Here, the rung interaction, the 1st neighbor interaction in a leg, and the 2nd neighbor interaction in a leg, are considered as the staggered 1st neighbor interaction, the 2nd neighbor interaction, and the 4th neighbor interaction in a chain, respectively. 
}
\label{fig:fsl}
\end{center}
\end{figure}

To decrease the numerical cost, the Hamiltonian should be rewritten by a matrix-product operator.
We use the following matrix-product kamiltonian:
\begin{equation}
\mcl{H}_\parallel+\mcl{H}_\perp=\bm{H}_1\mbf{H}_2\cdots\mbf{H}_{2N-1}(\bm{H}_{2N})^T
\end{equation}
where the local matrix operators are given by,
\begin{align}
\bm{H}_1&=\begin{pmatrix}
0 & \bm{M}_1 & \bm{P}_1 & \bm{Z}_1 & 1 
\end{pmatrix},\\
\mbf{H}_m&=\begin{pmatrix}
1 & & & & \\
\bm{p}_m^T & \mbf{1}^\prime & & & \\
\bm{m}_m^T & & \mbf{1}^\prime & & \\
\bm{z}_z^T & & & \mbf{1}^\prime & \\
0 & \bm{M}_m & \bm{P}_m & \bm{Z}_m & 1 
\end{pmatrix},\\
\bm{H}_{2N}&=\begin{pmatrix}
1 & \bm{p}_{2N} & \bm{m}_{2N} & \bm{z}_{2N} & 0
\end{pmatrix},
\end{align}
with the rectangular identity matrix 
\begin{equation}
\mbf{1}^\prime=
\begin{pmatrix}
1 & 0 & 0 & 0\\
0 & 1 & 0 & 0\\
0 & 0 & 1 & 0\\
\end{pmatrix}.
\end{equation}
Here, we define local operators in the upper leg ($j=1,2,\cdots,N$), 
\begin{align}
\bm{M}_{2j-1}&=\frac{1}{2}
\begin{pmatrix}
J_\perp S_{j,\rnu}^- & J_1 S_{j,\rnu}^- & 0 & J_2 S_{j,\rnu}^- 
\end{pmatrix},\\
\bm{P}_{2j-1}&=\frac{1}{2}
\begin{pmatrix}
J_\perp S_{j,\rnu}^+ & J_1 S_{j,\rnu}^+ & 0 & J_2 S_{j,\rnu}^+ 
\end{pmatrix},\\
\bm{Z}_{2j-1}&=
\begin{pmatrix}
J_\perp S_{j,\rnu}^z & J_1 S_{j,\rnu}^z & 0 & J_2 S_{j,\rnu}^z 
\end{pmatrix},
\end{align}
and local operators in the lower leg ($j=1,2,\cdots,N$),
\begin{align}
\bm{M}_{2j}&=\frac{1}{2}
\begin{pmatrix}
0 & J_1 S_{j,\rnl}^- & 0 & J_2 S_{j,\rnl}^- 
\end{pmatrix},\\
\bm{P}_{2j}&=\frac{1}{2}
\begin{pmatrix}
0 & J_1 S_{j,\rnl}^+ & 0 & J_2 S_{j,\rnl}^+ 
\end{pmatrix},\\
\bm{Z}_{2j}&=
\begin{pmatrix}
0 & J_1 S_{j,\rnl}^z & 0 & J_2 S_{j,\rnl}^z 
\end{pmatrix}.
\end{align}
Similarly, we also use other local-operator vectors,
\begin{align}
 \bm{p}_{2j-1\> (2j)}&=
\begin{pmatrix}
 S_{j,\rnu\> (\rnl)}^+ & 0 & 0 & 0
\end{pmatrix},\\
 \bm{m}_{2j-1\> (2j)}&=
\begin{pmatrix}
 S_{j,\rnu\> (\rnl)}^- & 0 & 0 & 0
\end{pmatrix},\\
 \bm{z}_{2j-1\> (2j)}&=
\begin{pmatrix}
 S_{j,\rnu\> (\rnl)}^z & 0 & 0 & 0
\end{pmatrix}.
\end{align}
In these vectors, each elements represents the connectivity to $n$th neighboring sites in one dimension (see Fig.~\ref{fig:fsl}), e.g., the 1st (2nd) element of $\bm{M}_{2j-1}$, that is $\frac{J_\perp}{2}S_{j,\rnu}^-$ ($\frac{J_1}{2}S_{j,\rnu}^-$), couples to $S_{j,\rnl}^+$ ($S_{j,\rnu}^+$) as the 1st (2nd) neighbor spin operator. 
Using this matrix-product operator, the matrix element $H^{\bm{\sigma},\bm{\sigma}^\prime}$ is rewritten by a matrix-product form
\begin{equation}
H^{\bm{\sigma},\bm{\sigma}^\prime}=\bm{H}_1^{\sigma_1,\sigma_1^\prime}\mbf{H}_2^{\sigma_2,\sigma_2^\prime}\cdots\mbf{H}_{2N-1}^{\sigma_{2N-1},\sigma_{2N-1}^\prime}(\bm{H}_{2N}^{\sigma_{2N},\sigma_{2N}^\prime})^T.
\end{equation}
In this calculation, we choose the Hilbert space with the constant magnetization $M$.  

\section{Extrapolation of system size}
\begin{figure*}[hbt]
\begin{center}
\includegraphics[keepaspectratio,width=0.9\textwidth]{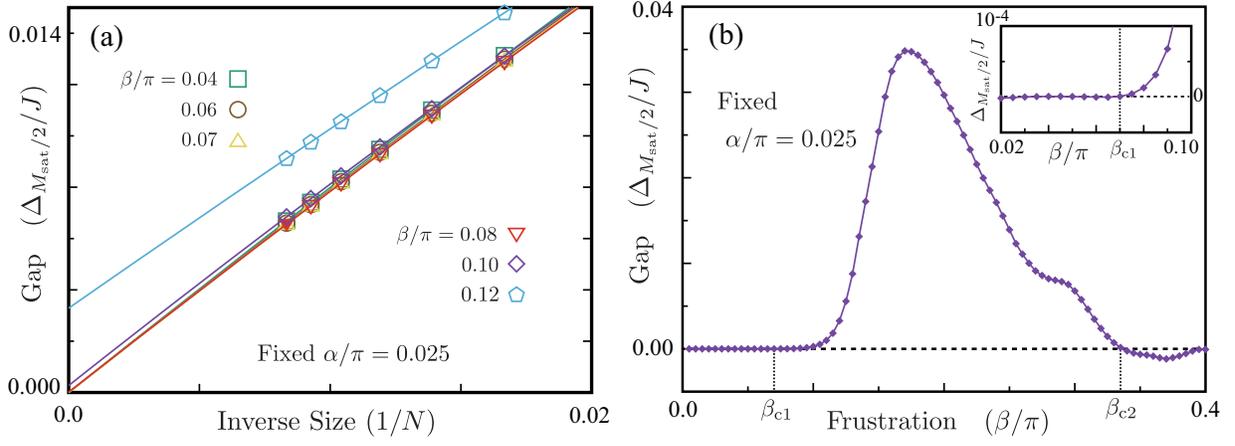}
	\caption{(a) Extrapolated MP gaps $\Delta_{M_{\mrm{sat}}/2}$ for various frustration $\beta=0.04$ to $0.12$ with a fixed leg-rung ratio $\alpha=0.025$. (b) Extrapolated MP gaps $\Delta_{M_{\mrm{sat}}/2}$ as a function of the frustration $\beta$. The inset shows an enlarged view around the critical frustration $\beta_{\mrm{c}1}$.
}
\label{fig4}
\end{center}
\end{figure*}

We performe the variational matrix-product state calculation to obtain the magnetization-plateau (MP) phase diagrams with number of kept states $m=1000$ at most.
In this calculation, we confirm that the truncation error is less than about $10^{-5}$.
To determine the phase boundaries of the MP states, we first calculate ground-state energies $E_M(N)$ with finite magnetization $M$ in finite system size $N$ for the frustrated spin ladder Eq.~(\ref{eq:ham}).
The fluctuation of energies originating from the truncation error is approximately estimated by a product of the energy and the truncation error, $E_M(N)\times 10^{-5}$.
The MP gaps $\Delta_M$ are obtained as extrapolated values of $\Delta_M(N)=E_{M+1}(N)+E_{M-1}(N)-2E_M(N)$ with respect to inverse system size $1/N\to0$ (see Fig.~\ref{fig4} (a)).

In this extrapolation, we use a second-order polynomial function $\Delta_{M_{\mrm{sat}/2}}(N)=aN^{-2}+bN^{-1}+\Delta_{M_{\mrm{sat}}/2}(N\to\infty)$ as a fitting function with constant values $a$, $b$, and $\Delta_{M_{\mrm{sat}}/2}(N\to\infty)$. 
We can see a good fitting and a transition from zero $\Delta_{M_{\mrm{sat}}/2}(N\to\infty)$ to finite as a result of fitting in Fig.~\ref{fig4} (a).

By using the extrapolation, we second plot the MP gap $\Delta_{M_{\mrm{sat}}/2}(N\to\infty)$ as a function of frustration in Fig.~\ref{fig4} (b).
This figure shows two transition point of $\beta$ with a fixed $\alpha=0.025$.
The lower critial point $\beta_{\mrm{c}1}$ is a transition from gapless to gapped, which gives a good coincidense with the Berezinskii-Kosterlitz-Thouless (BKT) transition expected in the first-order effective Hamitonian Eq.~(\ref{eq:ham1}).  
The upper critial point $\beta_{\mrm{c}2}$ seems to a transition from positive gap to negative one, which implies that $M_{\mrm{sat}}/2$ state is {\it unphysical} a certain region over $\beta_{\mrm{c}2}$. Namely, this state is skipped as increasing magnetic field, because a $M_{\mrm{sat}}/2+1$ state becomes less-energy state than any $M$ states before a $M$ state becomes the ground state.
This feature can originate from a finite binding energy of several triplons or another magnetic quasi-particles.

In the MP phase diagrams, error-bars represent tics of sampling points including the fluctuation energy estimated by the truncation error, i.e., the errror bars contains the regions with a comparable gap to the fluctuation energy.

\begin{thebibliography}{00}
\bibitem{note2} Regarding BiCu$_2$PO$_6$, we need to consider the Dzyaloshinskii-Moriya interaction, although most of our results will be satisfied qualitatively.
\end{thebibliography}
\end{document}